# Thermal Conductivity of GaAs/Ge Nanostructures

Roger Jia[1], Lingping Zeng[2], Gang Chen[2], and Eugene A. Fitzgerald[1]

[1]Department of Materials Science and Engineering, Massachusetts Institute of Technology, Cambridge, MA 02139, USA

[2]Department of Mechanical Engineering, Massachusetts Institute of Technology, Cambridge, MA 02139, USA

**Abstract**

Superlattices are promising low-dimensional nanomaterials for thermoelectric technology that is capable of directly converting low-grade heat energy to useful electrical power. In this work, the thermal conductivities of GaAs/Ge superlattice nanostructures were investigated systematically in relation to their morphologies and interfaces. Thermal conductivities were measured using ultrafast time-domain thermoreflectance and were found to decrease with increasing interface densities, consistent with our understanding of microscopic phonon transport in the particle regime. Lower thermal conductivities were observed in $(GaAs)_{0.77}(Ge_2)_{0.23}$ alloys; transmission electron microscopy study reveals phase separation in the alloys. These alloys can be interpreted as fine nanostructures, with length scales comparable to the periods of very thin superlattices. Our experimental findings help gain fundamental insight into nanoscale thermal transport in superlattices and are also useful for future improvement of thermoelectric performance using superlattice nanostructures.

**Introduction**

Approximately 60% of the energy consumption in United States is rejected as waste heat.[1] Thermoelectric energy conversion technology is a solid state technique that can directly convert low-grade waste heat into useful electrical power and has the potential to recover and reutilize the vast amount of rejected energy. The efficiency of a thermoelectric material is determined by the dimensionless thermoelectric figure of merit, $zT = \frac{\sigma S^2}{\kappa_e + \kappa_l}T$, where $\sigma$ is the electrical conductivity, $S$ is the Seebeck coefficient, $\kappa_e$ is the electronic thermal conductivity, $\kappa_l$ is the lattice thermal conductivity, and $T$ is the material's average operating temperature. Generally, a larger figure of merit yields a higher energy conversion efficiency. Desirable thermoelectric materials would therefore have high electrical conductivities and Seebeck coefficients but low thermal conductivities. Nanostructured materials, such as thin films, nanowires, and superlattices, have been extensively studied for thermoelectric applications over the past two decades due to their unique properties favorable for thermoelectrics. In particular, these materials typically exhibit superior thermoelectric figure of merit in comparison to their bulk constituents due to increased disruption to phonon transport at material interfaces and grain boundaries that significantly shortens the effective phonon mean free paths and thus leads to a significant reduction in the lattice thermal conductivity $\kappa_l$.[2–11]

Previous studies on thermal conductivity of superlattices have typically focused on homovalent structures such as GaAs/AlAs[12,13] and Si/Ge.[14,15] The fabrication of high quality superlattices of these materials is well established, allowing for high level investigations on specific phonon scattering processes. Heterovalent nanostructures, such as those containing both III-V and IV materials, can in principle produce more disruption to phonon transport by increasing the complexity of the material and thus increasing the number of scattering mechanisms. However, to the best of our knowledge, the thermal conductivities of these materials have not been studied in detail; only a few fabrication studies on molecular beam epitaxy (MBE) of GaAs/Ge superlattices, the simplest III-V/IV system, were conducted.[16–18] More recently, we investigated the fabrication of GaAs/Ge superlattices by metalorganic chemical vapor deposition (MOCVD) and observed crystal imperfections such as antiphase boundaries (APBs), layer roughness, and islands.[19]

Crystal imperfections are generally very difficult to account for when attempting to correlate high level physical processes to material structure. On a practical level, however, only the general effect of these imperfections on thermal conductivity is of interest. Past studies on GaAs/AlAs[20] and Si/Ge[21] superlattices have found that the cross-plane thermal conductivities are lower than the in-plane thermal conductivities. In addition, Luckyanova et al. observed that experimental results for thermal conductivity were lower than the results from simulations using density functional perturbation theory (DFPT), and suggested that this was a result of the simulations only accounting for atomistic mixing and not local layer thickness fluctuations.[20] These observations suggest that a traditional high quality material structure, as in a GaAs/AlAs superlattice, is not necessarily ideal for obtaining low thermal conductivities; the structural imperfections observed in the GaAs/Ge superlattices[10] thus may be beneficial for achieving lower thermal conductivities. This can be extended further to a nanostructure with randomized interfaces.

In this work, we systematically investigate the impact of sample morphologies and interface density on the thermal conductivity of GaAs/Ge nanostructures. We also study the thermal conductivity of a representative GaAs/Ge alloy structure in comparison to that of the GaAs/Ge nanostructures. The aim is to gain an understanding of the structure-property correlation of these GaAs/Ge structures and serve as a basis for further investigations on more complex III-V/IV nanostructures.

**Sample fabrication**

GaAs/Ge superlattice nanostructures were fabricated using MOCVD. Six samples consisting of 5, 8, 10, 30, 60, and 120 periods were fabricated by alternately depositing GaAs and Ge at a 50:50 GaAs:Ge composition. For clarity, we will use common superlattice terminology in discussing the experimental design and for sample identification; the actual observed structures are best described as nanostructures. Deposition times for the samples were adjusted to approximately maintain a total 500nm film thickness for each sample. An alloy sample was fabricated by simultaneously depositing both GaAs and Ge; the film thickness of this sample was 1μm. All films were grown at 650°C and at reactor pressures of 250 Torr. The substrates used for each sample were exact (100) GaAs substrates and (100) GaAs substrates offcut 6° toward the <111>A direction. For brevity we will refer to the samples as on-axis samples and offcut samples, respectively, when referencing the substrate tilt.

The morphology of the samples was analyzed in detail using cross-sectional transmission electron microscopy (XTEM); energy-dispersive X-ray spectroscopy (EDX) was used to estimate the composition in the alloy sample. The thermal conductivities of the samples were measured using an ultrafast time-domain thermoreflectance pump-probe setup at room temperature. A detailed discussion of the thermal conductivity measurement technique can be found in the relevant references.[22,23] Briefly, a thin (~100 nm) metal layer was coated onto the sample surface using electron beam deposition to act as an optical-thermal transducer in the measurement. The thickness of the metal film was accurately determined through TDTR measurement on a dummy sapphire substrate with known thermal conductivity that was co-placed in the deposition chamber with other GaAs/Ge superlattice samples. The thermal conductivities of the superlattice nanostructure samples were extracted by fitting the measured temporal thermoreflectance signal with the standard multi-layer heat diffusion model.

**Results and Discussion**

Images of the on-axis samples with 5, 30, 60, and 120 periods are shown in Fig. 1. Samples with lower number of periods, as in the 5 period, 8 period (not shown), and 10 period (not shown) samples, maintain alternating bands of GaAs and Ge. In the higher period samples, however, alternating GaAs and Ge bands was not achieved. This is very likely the result of the same challenges seen in previous studies – the preferential growth or bonding of GaAs to GaAs regions and Ge to Ge regions to avoid the thermodynamically unfavorable Ge-Ga and Ge-As bonds.[16,19] A consequence of reduced thicknesses of deposited layers would be an inability of GaAs, which nucleate as islands on Ge surfaces, to fully merge into a single layer. The final result could be a structure as seen in Fig. 1b, where GaAs islands are isolated by Ge and form a cell-like structure. In addition, near surface diffusion may play a role in the growth of the 120 period sample to reduce the expected number of interfaces. A comparison between the 60 period sample (Fig. 1c) and the 120 sample (Fig. 1d) indicates that the 120 sample appears to have approximately the same, if not less, number of interfaces.

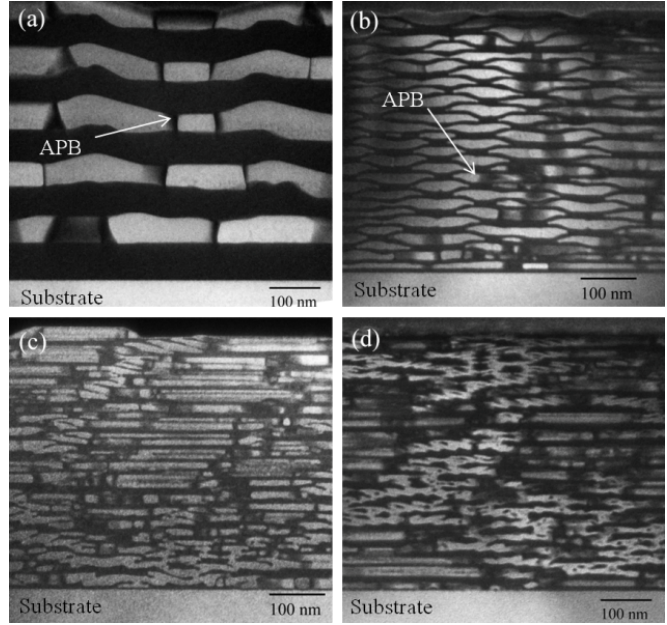

Figure 1. (002) Darkfield XTEM image of GaAs/Ge samples with (a) 5 periods, (b) 30 periods, (c) 60 periods, and (d) 120 periods. Bright bands/regions correspond to GaAs while dark bands/regions correspond to Ge. Antiphase boundaries (APBs) can be seen as dark vertical lines in the GaAs regions.

As discussed before, imperfections observed in the GaAs/Ge structures are not necessarily detrimental to achieving low thermal conductivities. For a comparison between samples, however, an interface density will need to be determined. The interface density of each sample was estimated by measuring a total interface length within an image and dividing by the encompassing nanostructure area. These interface densities were averaged over many images at various locations in each sample to reduce the measurement error. A representative trace of the measured reflectance signal along with the corresponding model fits is shown in Fig. 2, where the solid blue curve represents the best model fit from the heat diffusion theory and the dashed curves represent the model fits by adjusting the best fit thermal conductivity by ±10% to show measurement sensitivity. Generally, we achieved excellent fitting between the heat diffusion model and the experimentally measured reflectance signal by using the sample thermal conductivity and the metal-sample interface conductance as free parameters, indicating the adequacy of the thermal model. Figure 3 shows the correlation between the thermal conductivity and interface density of on-axis samples. The general trend of decreasing thermal conductivities as the interface density increases matches that of cross-plane thermal conductivities for GaAs/AlAs superlattices[13] and Si/Ge superlattices.[14] However, in contrast to those two systems, the lowest observed thermal conductivity corresponded to the alloy sample. Similar results were seen in the offcut samples, as shown in Fig. 4. A possible explanation for the low thermal conductivities of the alloy samples is phase separation of the sample into GaAs-rich and Ge-rich phases as a result of the immiscibility of GaAs and Ge. This could result in a heterogeneous alloy, with a finer nanostructure than that of the highest period samples and still maintain the phonon scattering mechanisms that allow for improved thermal conductivities over a homogeneous alloy.

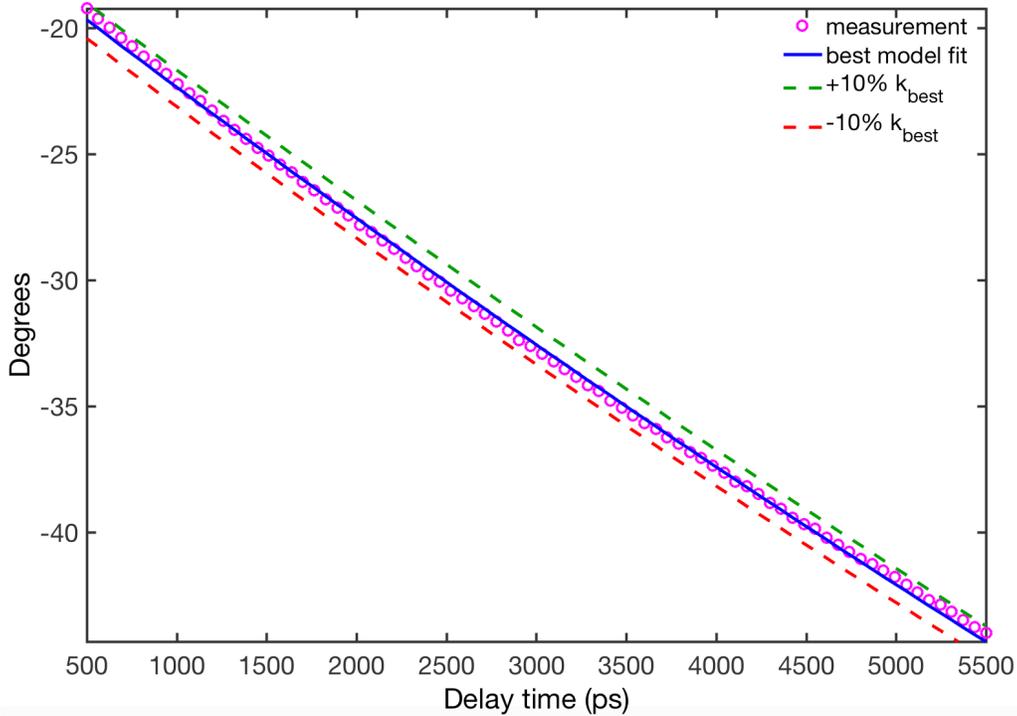

Figure 2. Representative measured phase signal along with the best model fit based on the thermal diffusion model. The dashed curves represent the model fits by adjusting the best fit

thermal conductivity by ±10% to show the sensitivity of the measurement to the underlying thermal conductivity.

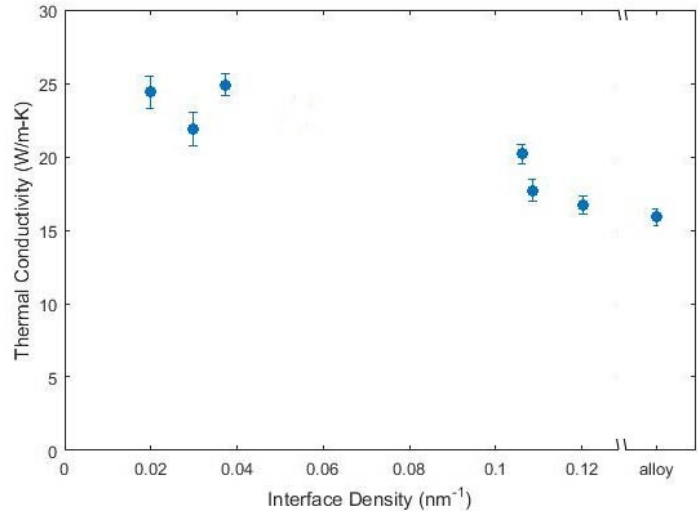

Figure 3. Thermal conductivities and the corresponding interface density of the on-axis samples. The alloy sample is shown on the right.

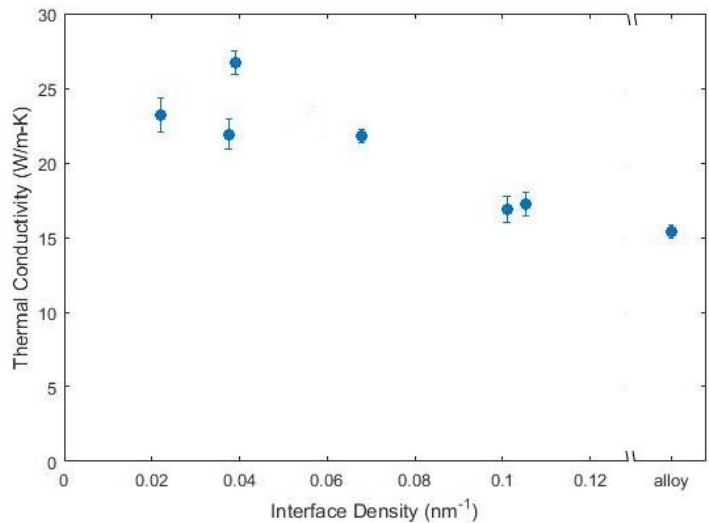

Figure 4. Thermal conductivities and the corresponding interface density of the offcut samples. The alloy sample is shown on the right.

The morphology of the offcut alloy sample, determined to be $(GaAs)_{0.77}(Ge_2)_{0.23}$, is shown in Fig. 5. The observed contrast in the alloy layer suggests that there is indeed some phase separation occurring. Interestingly, Norman et al. observed pronounced phase separation of $(GaAs)_{1-x}(Ge_2)_x$ alloys grown by MOCVD, forming a network of Ge-rich ribbons within a GaAs-rich matrix.[24] Their observed morphologies are comparable to the sample shown in Figure 1b, but not to the alloy sample in Fig. 5. It is possible that fabrication conditions play a role in this difference. Due to the low volume diffusion of Ge in GaAs and vice versa, pronounced phase separation would occur for a very prolonged growth process or by surface diffusion. If conditions were chosen such that both the fabrication process and film growth rates are fast, then phase separation could be significantly reduced.

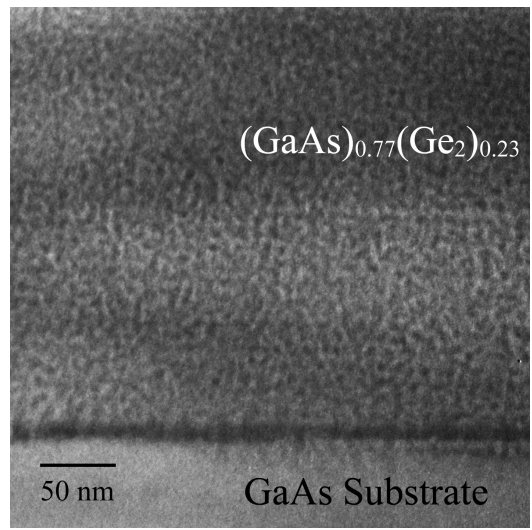

Figure 5. (002) Darkfield XTEM image of the $(GaAs)_{0.77}(Ge_2)_{0.23}$ alloy offcut sample.

**Conclusion**

We observed decreasing thermal conductivities in GaAs/Ge nanostructures with increasing interface density. The lowest thermal conductivities were seen in the $(GaAs)_{0.77}(Ge_2)_{0.23}$ alloys, in contrast to previous experimental studies done on GaAs/AlAs and Si/Ge superlattices. This is likely due to the fact that the $(GaAs)_{0.77}(Ge_2)_{0.23}$ alloys are heterogeneous in nature as a result of the immiscibility of GaAs and Ge. This also opens a direction for exploring more complex III-V/IV nanostructures such as InGaAs/SiGe, which would be expected to have significantly lower thermal conductivities. By fabricating as an alloy, potential issues that arise due to lattice mismatch with the substrate may possibly be avoided or reduced, and a wider range of compositions may be explored.

**Acknowledgements**

This work was supported under a research grant from the Singapore-MIT Alliance for Research and Technology Low Energy Electronic Systems Program, a research program funded by the National Research Foundation of Singapore. This work was also supported as part of the Solid State Solar Thermal Energy Conversion Center (S$^3$TEC), an Energy Frontier Research Center funded by the U.S. Department of Energy, Office of Science, Office of Basic Energy Sciences under award DE-SC0001299/DE-FG02-09ER46577. This work made use of the MRSEC Shared Experimental Facilities at MIT, supported by the National Science Foundation under award DMR-14-19807.

The authors would also like to thank Dr. Kimberlee Collins and Dr. Maria Luckyanova for advice and guidance on the thermal conductivity measurements.